\documentclass[twocolumn,showpacs,preprintnumbers,amsmath,amssymb]{revtex4}

\usepackage{graphicx}
\usepackage{dcolumn}
\usepackage{bm}
\input epsf

\begin{document}

\preprint{APS/123-QED}

\title{Multiplet Structure of Feshbach Resonances in non-zero partial waves}

\author{C. Ticknor, C. A. Regal, D. S. Jin\cite{byline}, J. L. Bohn}
\affiliation{JILA National Institute of Standards and Technology,
and Department of Physics,
University of Colorado, Boulder, CO 80309-0440}
\date{\today}

\begin{abstract}
We report a unique feature of magnetic field Feshbach resonances in which atoms
collide with non-zero orbital angular momentum.  P-wave ($l=1$) Feshbach 
resonances are split into two components 
depending on the magnitude of the resonant state's projection of orbital 
angular momentum onto the field axis.  This splitting is due to the magnetic 
dipole-dipole interaction between the atoms and it offers a means to tune 
anisotropic interactions of an ultra-cold gas of atoms.  A parameterization
of the resonance in terms of an effective range expansion is given. 
\end{abstract}

\pacs{Valid PACS appear here}
\maketitle
\section{Introduction}
The experimental observation of magnetic field Feshbach resonances (FRs) 
offers a means by which to widely tune the effective interactions in 
degenerate quantum gases.  A Feshbach resonance occurs when a quasibound 
state of two atoms becomes degenerate with the free atoms
and the interatomic potential either gains or loses a bound state.
As the quasibound state passes through threshold the scattering length can
be varied in principle from positive to negative infinity.  
FRs were observed in bosons \cite{Inouye,Courteille,Roberts1,Vuletic,marte},
in fermions between distinct spin states 
\cite{Loftus,Jochim,Ohara,Dieckmann}, and in a single-component Fermi gas
\cite{Regal}.  Using FRs to study fermions offers a means to explore 
superfluid phase transitions \cite{Stoof,Holland}, three-body recombination 
\cite{Esry}, mean-field interactions \cite{Regal2,Bourdel,Gupta} 
and molecules \cite{Regal3,Strecker,Cub,Jochim2}.

Of special interest is the p-wave FR observed in Ref. \cite{Regal}, which 
exists in a single-component Fermi gas.  Due to the Pauli exclusion principle 
the two-body wave function must be anti-symmetric under interchange of two 
fermions, implying that only odd partial waves $l$ can exist for identical 
fermions.  For $l=1$ the Wigner threshold law dictates that the p-wave cross 
section scales as the temperature squared.  This characteristic behavior 
ordinarily suppresses interactions
at ultra-cold temperatures\cite{Demarco}.  However, a resonance can 
dramatically increase the p-wave cross section even at low temperatures.

In this article we discuss characteristics of p-wave 
Feshbach resonances.  The first is a
sensitive dependence of observables on temperature and magnetic field. 
This dependence arises from a centrifugal barrier through which the wave 
function must tunnel to access the resonant state. 
Only in a narrow range of magnetic field can the continuum 
wave function be significantly influenced by the bound state.

The second characteristic is a doublet in the resonance feature arising from 
the magnetic dipole-dipole interaction between the atoms' valence electrons. 
The p-wave FRs experience a non-vanishing dipole-dipole interaction in lowest 
order, in contrast to s-wave FRs.  This interaction splits the FR into 
distinct resonances based on their partial wave projection onto the field axis,
$m_l=0$ or $|m_l|=1$.  Splitting of 
the p-wave resonance offers a means to tune the anisotropy of the 
interaction.  Dipole-dipole interactions in Bose-Einstein condensates and 
degenerate Fermi gas have been considered due to the novelty of the resulting 
strong anisotropic interaction \cite{polar}.  P-wave FRs offer an immediately 
accessible means to explore anisotropic interactions in the many-body physics 
of degenerate gases. 

The observed p-wave FR in  $^{40}$K occurs when two atoms in the
$|f,m_f \rangle$ $=|9/2, -7/2 \rangle$ 
hyperfine state collide. The joint state of the atom pair 
will be written 
$|f_1 m_{f_1} \rangle|f_2 m_{f_2} \rangle|lm_l\rangle=
|9/2, -7/2 \rangle |9/2, -7/2 \rangle|1,m_l \rangle$ 
where $m_l$ can take on the values $\pm1,0$. The calculations presented here 
were performed using Johnson's log-derivative propagator method \cite{Johnson} 
in the magnetic field dressed hyperfine basis \cite{Burke2}. 
The potassium singlet and triplet potentials \cite{triplet,singlet}
are matched to long range dispersion potentials with $C_6=3927$ a.u.
and are fine tuned to yield the scattering lengths $a_s=104.0$ a.u.,
$a_t=174$ a.u. respectively.  With these values we are able
to reproduce the FRs measured in Ref. \cite{Loftus,Regal}.

\section{Temperature and Magnetic field dependence}
\begin{figure}
\centerline{\epsfxsize=7.0cm\epsfysize=7.0cm\epsfbox{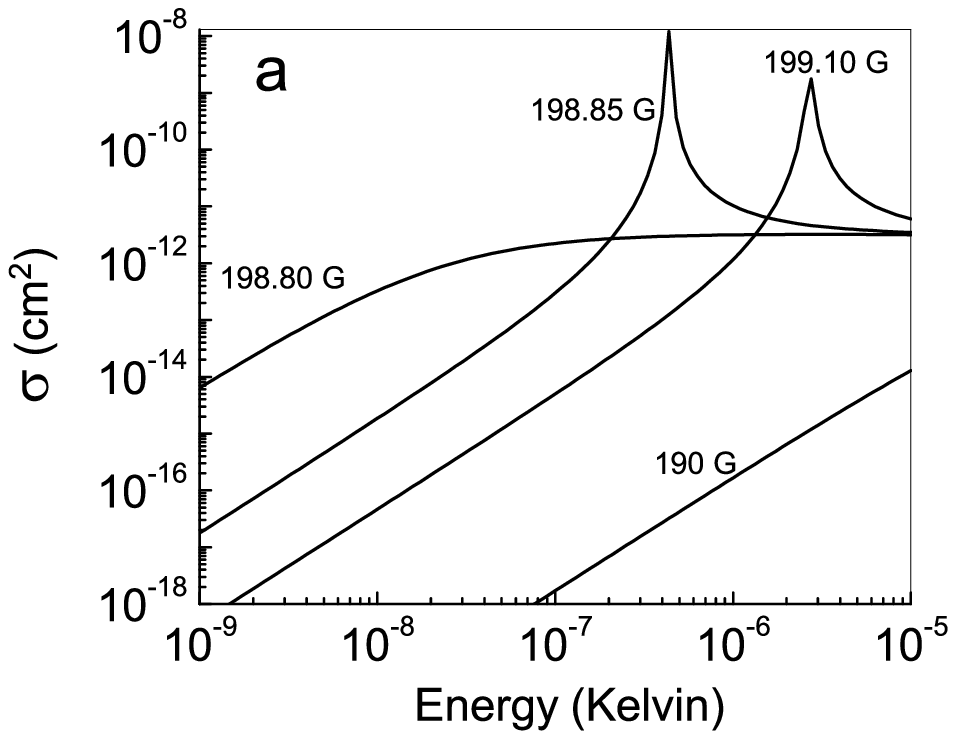}}
\centerline{\epsfxsize=7.0cm\epsfysize=7.0cm\epsfbox{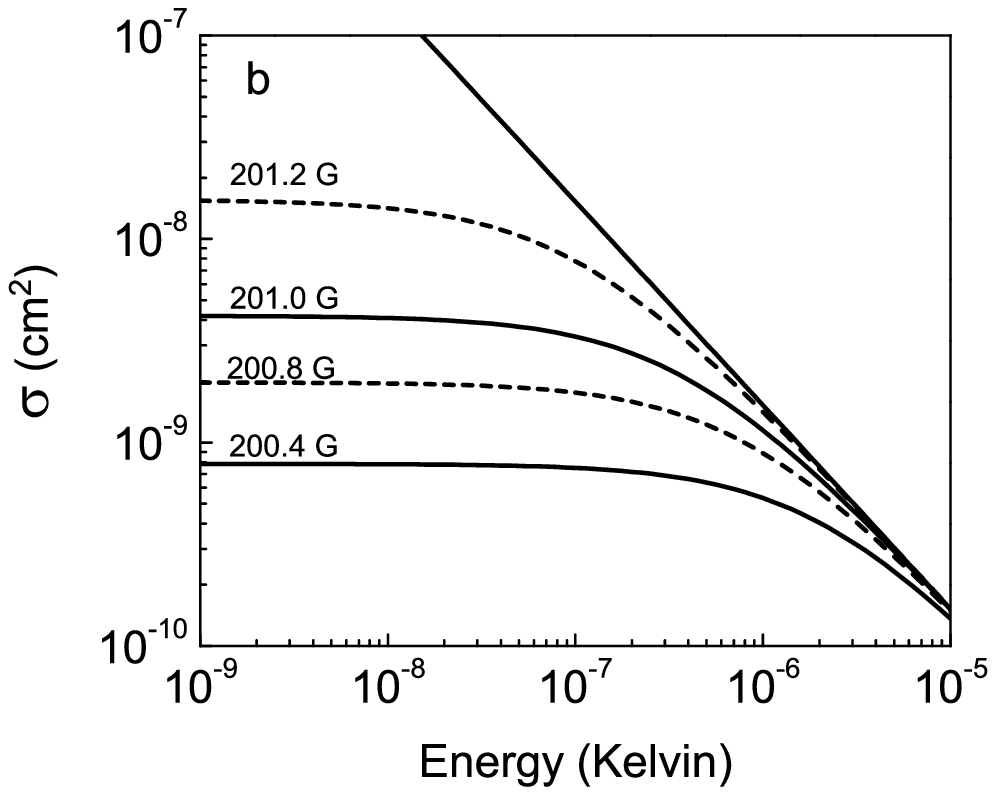}}
\caption{(a) P-wave elastic cross section versus energy for
$|9/2,-7/2\rangle|9/2,-7/2\rangle|1,1\rangle$ collisions for different magnetic
field values.  For each curve the magnetic field in gauss is indicated. 
The lowest curve shows an off-resonance cross section.
(b) For comparison, the s-wave elastic cross section versus energy for
$|9/2,-9/2\rangle|9/2,-7/2\rangle|0,0\rangle$ collisions for different magnetic
field values. The s-wave FR peaks at B=201.6 G.
Compared to the p-wave FR these have little structure.  The solid line 
is the unitarity limit.}
\label{energyde}
\end{figure}

A p-wave resonance is distinct from an s-wave ($l=0$) resonance
in that the atoms must overcome a centrifugal barrier to
couple to the bound state. 
The extreme dependence on magnetic field and temperature can be 
understood by considering the cross section as a function of energy at
several values of magnetic field, as shown in Fig. \ref{energyde} (a). 
The lowest curve, with a magnetic field B=190 G, shows typical 
off-resonance p-wave threshold behavior.
Once the magnetic field is increased to be close to the resonance, the 
cross section changes significantly, and a narrow resonance appears 
at low energy.  The resonance first appears for fields just above B=198.8 G.
As the magnetic field is increased the resonance 
broadens and moves to higher energy.  The p-wave resonance's narrowness  
is due the fact that atoms must tunnel through a centrifugal barrier before 
they can interact.
 
This narrow resonance structure is in stark contrast to the s-wave FR
shown in Fig. \ref{energyde} (b), which occurs between the spin states
$|9/2,-9/2 \rangle $ and $|9/2,-7/2 \rangle$, as reported in Ref. 
\cite{Loftus}.  The energy dependence of the s-wave FR has a much 
simpler form than the p-wave FR.
At high energy the cross section is essentially at the unitarity limit, which 
is shown as the solid line.
At lower energy the cross section plateaus at a constant value of $4\pi a^2$, 
where $a$ is the s-wave scattering length. 
The energy at which the cross section plateaus depends on the magnetic field. 
The closer to resonance the magnetic field is tuned, the lower the energy 
at which the cross section plateaus. 
 
The temperature dependence of p-wave FRs results from the strong energy 
dependence of the cross section.  For a Maxwell-Boltzmann distribution of the 
atomic energies, the thermally averaged cross section is
\begin{equation}
\langle \sigma \rangle={1\over (kT)^2} \int_0 ^\infty \sigma (E)E 
e^{-E/kT} dE
\end{equation}
where $k$ is Boltzmann's constant and $\sigma(E)$ is the energy dependent
cross section \cite{Burke2}.

\begin{figure}

\centerline{\epsfxsize=7.0cm\epsfysize=7.0cm\epsfbox{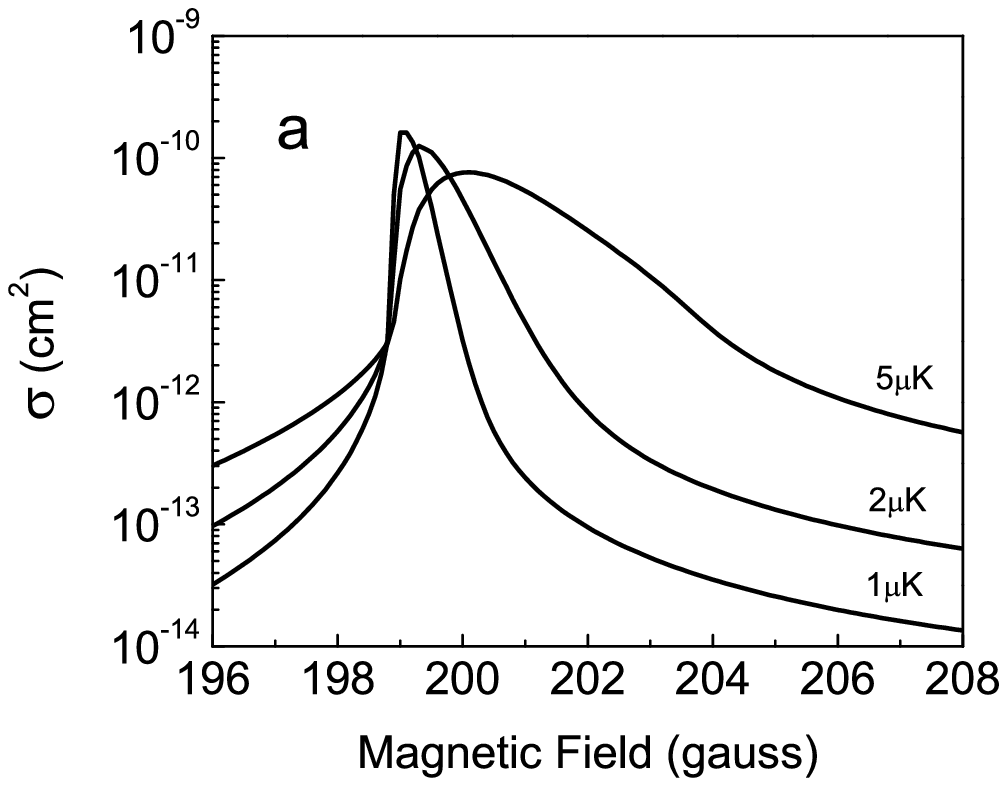}}
\centerline{\epsfxsize=7.0cm\epsfysize=7.0cm\epsfbox{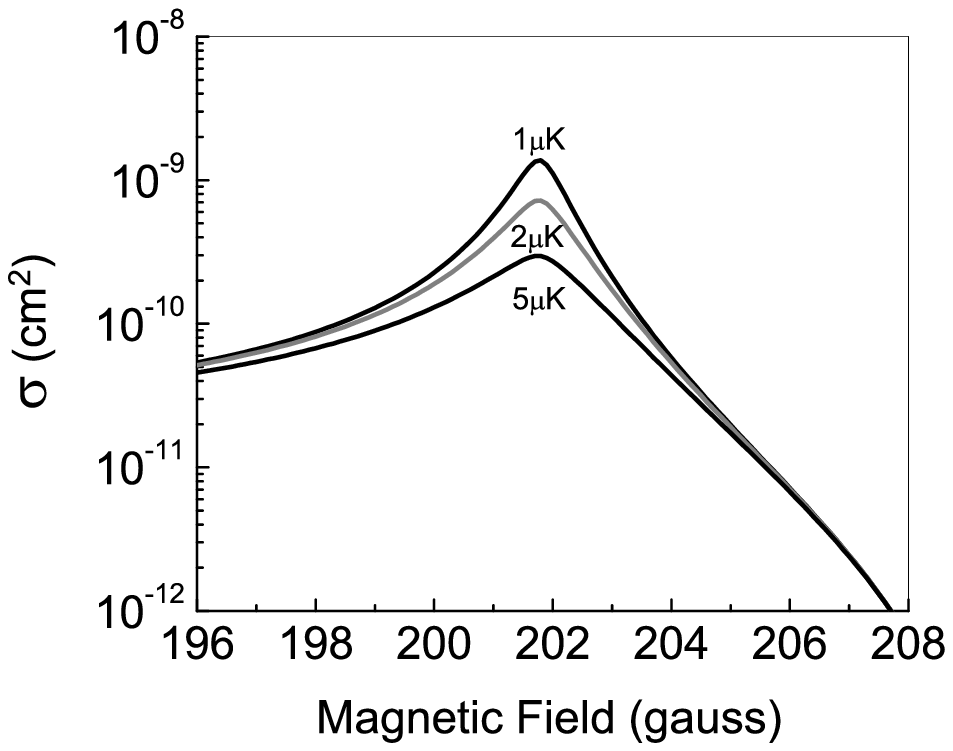}}
\caption{(a) Thermally averaged cross section for 
$|9/2,-7/2\rangle|9/2,-7/2\rangle|1,1\rangle$ collisions
as a function of magnetic field. 
The striking features of this curve are the sudden rise and change in 
width as the temperature is increased.
(b) Thermally averaged cross section for 
$|9/2,-9/2\rangle|9/2,-7/2\rangle|0,0\rangle$ collisions
as a function of magnetic field. The
temperature dependence is only evident at the peak where it washes out the
maximum value.}
\label{tempdep}
\end{figure}

Figure \ref{tempdep} shows the thermally averaged cross section for the
$m_l=1$ p-wave resonance, Fig. \ref{tempdep} (a), and for the s-wave FR,
Fig. \ref{tempdep} (b).
The key features of Figure \ref{tempdep} (a) are the sudden rise of the cross 
section and thermal broadening, which grows dramatically at high field as the 
temperature increases.  The rise comes from the sudden appearance of the 
the narrow resonance at positive collision energies as the magnetic field is 
tuned.  This rise is not temperature dependent because, regardless of
temperature, the threshold is first degenerate with the bound state at a 
unique magnetic field. By contrast, the high field tail of the resonance 
is sensitive to temperature because 
once the bound state has passed through threshold the energy dependent cross
section peaks at higher energies for higher field values.
For a fixed magnetic field the high field side of the FR, there is a well 
defined, narrow resonance at a particular energy 
(Fig. \ref{energyde} (a)).  If the temperature is low, very few atom
pairs can access this resonance.  At higher temperatures more atoms experience 
resonant scattering, increasing $\langle \sigma \rangle$.

This characteristic asymmetric profile is not present is s-wave FRs.  
The s-wave FR is shown in Fig. \ref{tempdep} (b) near its peak.  The only 
effect of temperature in the elastic cross section is to wash out the peak 
of the resonance as the temperature is increased.  This behavior follows from 
the relatively structureless energy-dependent cross section in Fig. 
\ref{energyde} (b).
\section{The Doublet Feature}
The valence electrons of ultra-cold alkali atoms interact
via a magnetic dipole-dipole interaction of the form
\begin{equation}
 H_{ss}=-\alpha^2
{3 ({\bf \hat R} \cdot {\bf \hat s}_1)({\bf \hat R} 
\cdot {\bf \hat s}_2)-{\bf \hat s}_1 \cdot {\bf \hat s}_2 \over R^3}
\label{fulldidi}
\end{equation}
where $\alpha$ is the fine structure constant, ${\bf \hat s}_i$ the spin of 
the valence electron on atom $i$, $R$ is the interatomic
separation, and ${\bf \hat R}$ is the normal vector defining the 
interatomic axis.  Another way of writing this interaction that 
isolates the spin and partial wave operators is 
\begin{equation}
\label{didi}
 H_{ss}= -{\alpha^2\sqrt{6}\over R^3} \sum_{q=-2}^2 (-1)^q 
C^2_q \cdot (s_1 \otimes  s_2)^2_{-q} .
\end{equation}
Here $C^2_q(\theta,\phi)$ is a reduced spherical harmonic
that depends on the relative orientation of the atoms, and 
$(s_1 \otimes  s_2)^2_{-q}$  
is the second rank tensor formed from the rank-1 spin operators \cite{Brink}.
$C^2_q$ acts on the partial wave component of the quantum state,
$|l m_l \rangle$, while the $s_i$'s in $(s_1 \otimes  s_2)^2_{-q}$ act on the 
electronic spin state of the atoms.
Equation (\ref{didi}) leads to an interplay of partial wave and spin, which
contributes an orientation-dependent energy to the Hamiltonian.
The matrix element of Eq. (\ref{didi}) in our present basis is \cite{Burke2}
\begin{eqnarray}
\label{ddbasis}
-{\alpha^2 \sqrt{6} \over R^3} \sum_{q=-2}^2 (-1)^q 
{\langle l^\prime m_l^\prime |C^2_q|l m_l \rangle}\times
\nonumber\\
{\langle f_1^\prime m_{f_1}^\prime| \langle f_2^\prime m_{f_2}^\prime
|(s_1 \otimes  s_2)^2_{-q}|f_1 m_{f_1} \rangle|f_2 m_{f_2} \rangle}.
\end{eqnarray}
This term in the Hamiltonian couples different partial waves for
$l^\prime=l\pm 2$, and it couples different partial wave 
projections $m_l$ for $l^\prime=l\pm 2$ and $l=l^\prime\neq 0$.  
For elastic s-wave scattering ($l=l^\prime=0$)
equation (\ref{ddbasis}) vanishes by symmetry. 
This term only plays a role in s-wave scattering for 
s$\rightarrow$d wave transitions.
However for p-wave scattering ($l=l^\prime =1$) this interaction 
does not vanish, i.e.  $\langle 1 m_l^\prime|C_q^2|1 m_l\rangle\neq 0$. 
Furthermore, for elastic scattering, $q=0$, the interaction depends on $m_l$, 
since $\langle 1 1|C_0^2|1 1\rangle=-{1\over5}$ and  
$\langle 1 0|C_0^2|1 0\rangle={2\over5}$.
The fact that the dipole-dipole interaction does not contribute 
equally to all values of $m_l$ means that bound states with different $m_l$
have different energies.
This implies that FRs with different values of $m_l$ couple to distinct 
bound states and thus have different magnetic field dependences.
   
The difference between the $m_l$ projections can be understood intuitively
by considering the dipole-dipole interaction of the two atoms.  
For the $|9/2,-7/2\rangle$ spin states case the electronic spins are 
essentially aligned with the field.  When two dipoles are aligned head to 
tail they are in an attractive configuration, corresponding to 
${\bf \hat R} \cdot {\bf \hat s_i}=1$ in Eq. (\ref{fulldidi}).
Vice-versa when the dipoles are side by side
they are in a repulsive configuration, ${\bf \hat R} \cdot {\bf \hat s_i}=0$. 

\begin{figure}
\centerline{
\epsfxsize=7.0cm\epsfysize=5.0cm\epsfbox{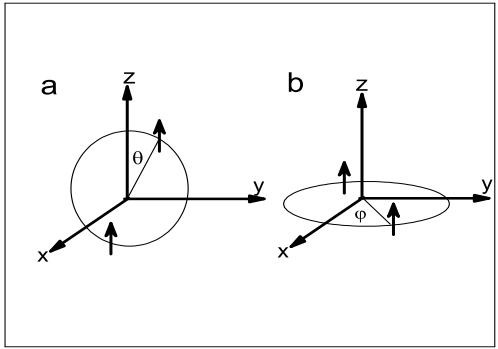}}
\caption{Schematic representation of classical dipoles interacting 
in different circular orbits.  Shown in  (a) is an orbit
with $m_l=0$, which is in a plane containing the magnetic field.   
Here the dipoles sometimes attract and sometimes repel.
In (b) is shown an orbit with $|m_l|=1$, in a plane perpendicular to the 
magnetic field.  Here the atoms predominately repel one another.}
\label{dischem}
\end{figure}

Viewing the motion of the atoms in the resonant state as classical, circular 
orbits, the cases $m_l=0$ and  $m_l=1$ are distinguished as in Figure 
\ref{dischem}.  For $m_l=0$, in Fig. \ref{dischem} (a), the motion of the 
atoms is in a plane containing the magnetic field.  Classically this 
corresponds to motion described by the angle $\theta$, where the magnetic 
field lies in the $\hat z$ direction.  The interaction for $m_l=0$ alternates 
between attractive and repulsive as the dipoles
change between head to tail attraction and side by side repulsion.  
On the other hand, for $|m_l|=1$, shown in Fig. \ref{dischem} (b),  
the motion 
of the atoms is in the plane  perpendicular to the magnetic field. Classically 
this corresponds to motion described by the angle $\phi$.  This interaction
is only repulsive, because the dipoles are held in the side-by-side 
configuration.  Since the dipole-dipole interaction for $|m_l|=1$ has only a 
repulsive influence it forms a resonant state with higher energy. 

\begin{figure}
\centerline{
\epsfxsize=7.0cm\epsfysize=7.0cm\epsfbox{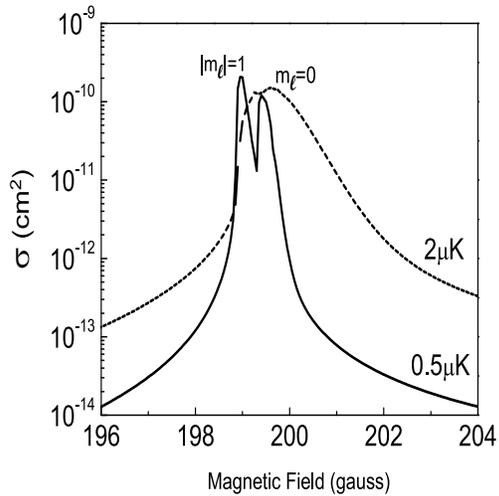}}
\caption{The thermally averaged elastic cross section for the p-wave 
FR, including all partial wave projections $m_l=-1,0,1$.  
At low temperatures, the doublet splitting emerges clearly, but it is washed 
out a higher temperatures due to thermal broadening.  The lower field 
resonance has $|m_l|=1$ and the higher field resonance has $m_l=0$.}
\label{fullsigma}
\end{figure}

Figure \ref{fullsigma} shows the total elastic cross section as a function of 
field at several temperatures.  One can clearly see the doublet feature in the 
cross section at low temperature.  The first peak corresponds to $|m_l|=1$.  
The doublet cannot be 
resolved at high temperature because the width of the resonance is wider 
than the splitting.

The energy shift can be estimated using perturbation theory.
The energy shift due to the dipole-dipole interaction is 
given in  perturbation theory as
\begin{eqnarray}
\label{pert1}
\Delta E_{m_l=0}={-\alpha^2 \surd6} \langle 1 0|C_0^2|1 0\rangle 
\langle \Phi_{mol}|{(s_1 \otimes  s_2)^2_{0}\over R^3}|\Phi_{mol}\rangle
\\
\label{pert2}
\Delta E _{m_l=1}={-\alpha^2\surd6 } \langle 1 1|C_0^2|1 1\rangle 
\langle \Phi_{mol}| {(s_1 \otimes  s_2)^2_{0}\over R^3}|\Phi_{mol}\rangle.
\end{eqnarray}
Here $|\Phi_{mol}\rangle$ is the full multi-channel molecular wave function 
without the magnetic dipole-dipole interaction.  This is the molecular
state that couples to the continuum creating the FR.  We notice 
that the perturbation is the same for each component, except for the
angular coefficients in Equations (\ref{pert1}) and (\ref{pert2}).  When 
these equations are evaluated, we find that the energy 
difference in the molecules,$|\Delta E_{m_l=1}-\Delta E_{m_l=0}|$, 
is $3.7 \mu$K, which is close to the 
closed coupling calculation result of $4.7 \mu$K.  As the bound states are 
brought through threshold, we find that their energy difference 
translates into a peak separation of $0.5$ G, determined from the closed 
coupling scattering calculations.

\begin{figure}
\centerline{
\epsfxsize=7.0cm\epsfysize=6.50cm\epsfbox{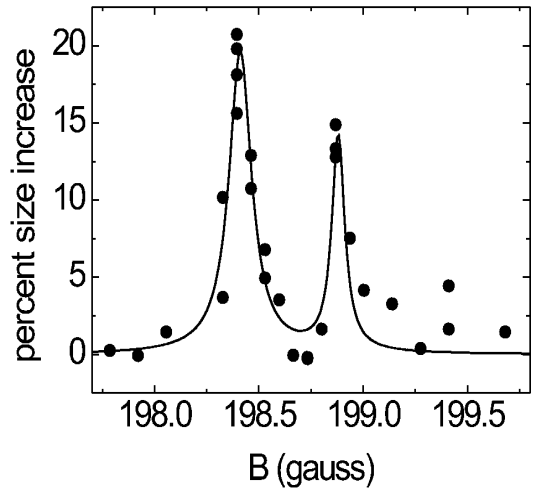}}
\caption{The p-wave FR observed through heating of the gas,
clearly showing the doublet feature of the p-wave resonance.  The 
cloud started at T=0.34 $\mu$K and then was held at a constant 
magnetic field. Inelastic processes at the FR, 3-body dominated, 
heat the cloud resulting in an increase in the measured size of the trapped 
cloud.  The curve is only a guide to the eye.}
\label{exp}
\end{figure}

Experimentally we have observed the doublet p-wave resonance in an 
ultra-cold gas of  $^{40}$K through inelastic collisional effects.  A gas of
atoms in the $|9/2,-7/2\rangle$ state of $^{40}$K was prepared at T=0.34 
$\mu$K in an optical dipole trap characterized by a radial frequency
of $\nu_r$=430 Hz and an axial frequency of  $\nu_z$=7 Hz \cite{Loftus,Regal}.
The gas was then held at a magnetic field near resonance for
260 $\mu$s.  The resulting Gaussian size of the trapped gas in the axial 
direction was measured as a function of magnetic field.  The result of this 
measurement is shown in Fig. \ref{exp}.

The observed heating of the gas in Figure \ref{exp} is due to inelastic 
processes that occur at the p-wave FR.
The result clearly shows the predicted doublet structure.  The splitting 
between the two peaks is measured to be 0.47$\pm$8 G, in good agreement with 
the theory. The dominant inelastic processes are 3-body losses 
\cite{Regal,Esry}, which lie at a slightly lower field than the elastic 
resonance peak \cite{Regal}.
\section{Effective Range Expansion of the p-wave FR}
To compute many-body properties of degenerate gases the s-wave scattering 
length is often used to mimic the essential 2-body physics.
Near a FR the scattering length diverges and can be represented well by
$ a=a_{bg}(1-{\Delta\over B-B_0})$ where 
$a_{bg}$ is the back ground scattering length, $\Delta$ is the width,
$B_0$ is the location of the s-wave resonance. Scattering length is defined as
$a=^{\lim} _{k \rightarrow 0}-\tan(\delta_0)/k$ 
where $\delta_0$ is the s-wave phase shift and 
$k=\sqrt{2\mu E}$ where $\mu$ is the reduced mass. 
 
\begin{figure}
\centerline{\epsfxsize=7.0cm\epsfysize=7.0cm\epsfbox{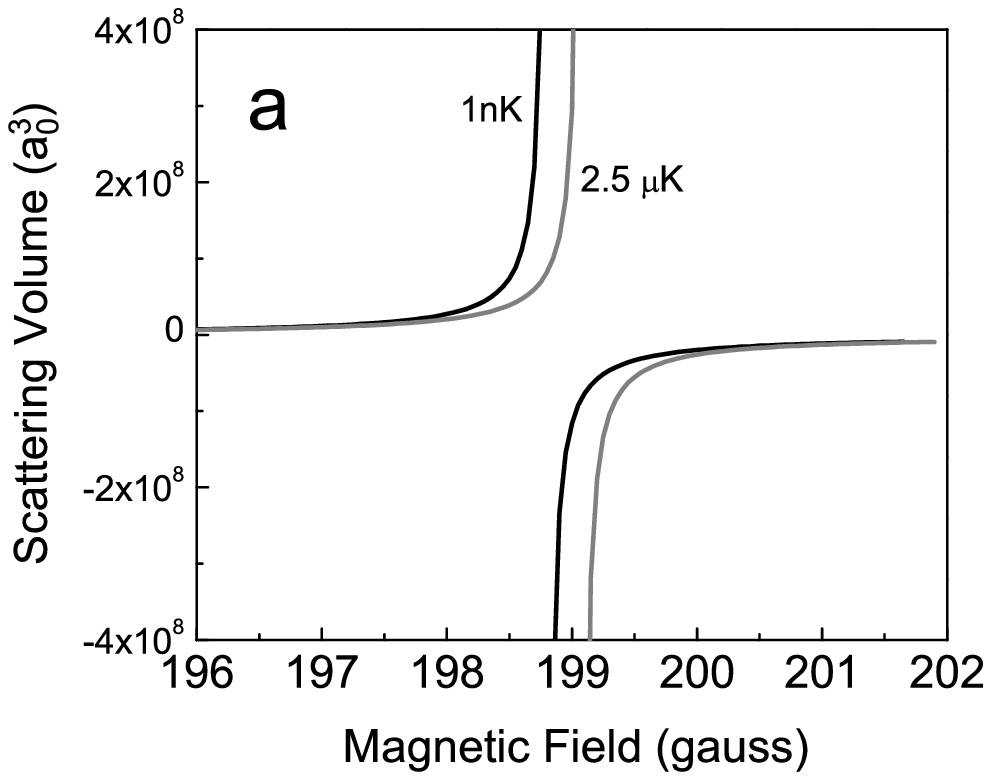}}
\centerline{\epsfxsize=7.0cm\epsfysize=7.0cm\epsfbox{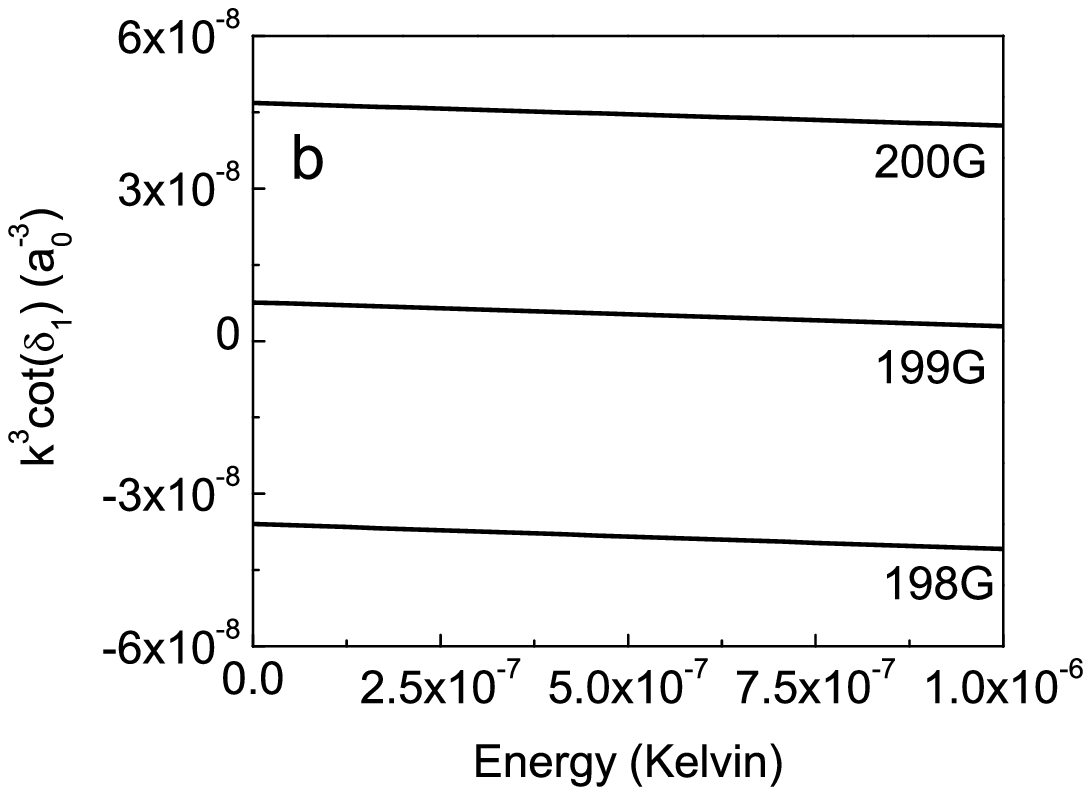}}
\caption{(a) The p-wave scattering volume 
for $|m_l|=1$ as a function
of magnetic field at two different energies.  Notice that both the location
at which the scattering volume diverges and the width vary 
with collision energy. (b) $k^3\cot(\delta_1)$ for
the $|m_l|=1$ p-wave resonance as a function of energy for
several different values of magnetic field.}
\label{scvol}
\end{figure}

For p-wave collisions the relevant quantity is the scattering volume,
$v=^{\lim} _{k \rightarrow 0} -\tan(\delta_1)/k^3$.  A simple form like the 
one 
for $a$ is inadequate for parameterization of the p-wave scattering volume
because the p-wave resonance has a complicated energy dependence.
Fig. \ref{scvol}(a)
shows the p-wave scattering volume as a function of field
and energy.  The curves show that as the energy is increased the location and
width of the resonance change. 
 
To adequately parameterize the p-wave phase shift across the resonance
one must use the second order term in the the effective range expansion 
\cite{MM}.
Fig. \ref{scvol}(b) shows $k^3\cot(\delta_1)$ plotted as a function of 
energy for several magnetic fields.  This set of curves can then 
be fit using effective range expansion of the form
\begin{equation}
k^3\cot(\delta_1)=-{1\over v}+{c}k^2.
\label{vsc}
\end{equation}
Where $\delta_1$ is the p-wave phase shift, $v$ is the scattering volume, 
and $c$ the second coefficient in the expansion, analogous to the  
effective range in the s-wave expansion, but with units $a_0^{-1}$. 
Both $v$ and $c$ are functions of magnetic field.
Fitting $v$ and $c$ to quadratic functions of $B$, which is adequate
for the energy range of $E < 10^{-6}$ K and magnetic field range 
of 195 to 205 gauss, we find
\begin{eqnarray}
\nonumber
{1\over v_{m_l=0}}=8.68155\times10^{-5} -8.29778\times10^{-7} B \\
\nonumber
+1.97732\times10^{-9} B^2\\
\nonumber
c_{m_l=0}=-1.64805 +0.01523 B-3.54471\times10^{-5} B^2\\
\nonumber
{1\over v_{|m_l|=1}}=7.83424\times10^{-5} -7.456621\times10^{-7} B \\
\nonumber
+ 1.76807\times10^{-9} B^2\\
\nonumber
c_{|m_l|=1}=-2.36792+0.02264 B-5.45051\times10^{-5} B^2
\end{eqnarray}
where B is magnetic field in gauss.  These fits for $1\over v$
and $c$ accurately reproduce $k^3\cot(\delta_1)$
to within $ 3 \% $ on the interval specified.  This fit does not include
experimental uncertainties, rather the fit is designed to reproduce the
the closed coupling calculation with the optimal scattering parameters.
\section{Implications for experiments}
Because of the p-wave FR, angular dependence of scattering is
under the experimenter's control.  For example if magnetic
field tuned to 199.0 G in the $^{40}K$ p-wave FR the dominant interactions 
in the gas will be perpendicular to the field axis, which follows from the
angular distribution corresponding to the spherical harmonic 
$Y_{11}(\theta,\phi)$ leading to enhanced collisions rates away from the field 
axis.  Whereas if the field is tuned 0.5 G higher, at cold 
temperatures the interaction will be dominated by $Y_{10}(\theta,\phi)$,
characterized by enhanced collisions along the field axis.
The angular dependence of the collisions also has implications for
the inelastic 2-body processes.  These processes are characterized by
two atoms gaining a predictable amount of energy governed by hyperfine 
splitting and are redistributed in a well defined angular manner.

P-wave FRs offer a means to experimentally study anisotropic interactions
in systems other than identical fermions.  For example, we predict that 
there are p-wave FRs in distinct spin states of bosonic $^{85}$Rb and in 
the Bose-Fermi mixture of $^{87}$Rb-$^{40}$K, shown in Table \ref{rb}. 
The Rb calculations used potentials that are consistent with Ref. 
\cite{Roberts}.  The  Rb-K  calculations are consistent with Ref. 
\cite{Simoni}.  On resonance the p-wave cross section becomes comparable 
to the background s-wave scattering.  This means that it could have an 
equally important role in determining the collisional behavior and mean-field 
interaction of a thermal gas or condensate.   
\section{Conclusion}
We have presented characteristics of p-wave FRs.  An interesting characteristic
is the doublet feature for the FR with $l=1$.  The splitting is caused by the 
dipole-dipole interaction having distinct values depending on partial 
wave projection.  This might in turn offer a means to study anisotropic 
interactions in quantum gases. Another feature of the p-wave FR is the 
asymmetric thermal broadening, which arises from the 
resonant state moving away from threshold as the magnetic field is tuned.
Generally, we expect the broadening to occur for fields which the $v<0$.
Just as a p-wave FR splits into two components  an $l$-wave
FR will split into $l+1$ components.
\begin{acknowledgments}
This work was supported by NSF, ONR, and NIST.
C.A.R. acknowledges support from the Hertz foundation. 
\end{acknowledgments}
\begin{table}
\caption{Predicted p-wave resonances in bosonic Rb and a Bose-Fermi mixture 
of Rb-K.  The Rb p-wave FRs gain a bound state as the field is increased, 
whereas the opposite is true for $^{40}K$ p-wave FR.  This is why in Rb the 
${m_l=0}$ resonance is lower in field.}
\begin{tabular}{cccc}
Species & Spin States & Magnetic Field &\\
\tableline
$^{85}Rb$ & $|2,-2 \rangle |2, -1 \rangle$&  
$B_{m_l=0}=247.3$ &  $\pm$ 5 G\\ 
 & & $B_{|m_l|=1}=248.0$ & $\pm$ 5 G\\
\tableline
$^{85}Rb$ $^{87}Rb$&$|2,-2\rangle_{85}|1,-1\rangle_{87}$ 
& $B_{m_l=0}=292.8$ & $\pm$ 30 G\\
 & & $B_{|m_l|=1}=292.5$ & $\pm$ 30 G\\
\tableline
$^{87}Rb$ $^{40}K$&$|1,1\rangle|9/2,-9/2\rangle$
& $B_{m_l=0}=540.0$ & $\pm$ 30 G\\
 & & $B_{|m_l|=1}=540.3$ & $\pm$ 30 G\\
\tableline
\end{tabular}
\label{rb}
\end{table}

\newpage

\begin{thebibliography}{99}
\bibitem[*]{byline}Quantum Physics Division, 
National Institute for Standards and Technology.
\bibitem{Inouye} S. Inouye, M. R. Andrews, J. Stenger, H.J . Miesner, D. M.
Stamper-Kurn, and W. Ketterle, Nature (London) {\bf 392}, 151 (1998).
\bibitem{Courteille} Ph. Courteille, R. S. Freeland, D. J. Heinzen, 
F.A. van Abeelen, and B. J. Verhaar, Phys. Rev. Lett. {\bf 81}, 69 (1998).
\bibitem{Roberts1} J. L. Roberts, N. R. Claussen, J.P. Burke, C. H. Greene, 
E. A. Cornell, and C. E. Wieman Phys. Rev. Lett. {\bf 81}, 5109 (1998).
\bibitem{Vuletic} V. Vuleti\'c, A.J. Kerman, C. Chin, and S. Chu, Phys.
Rev. Lett. {\bf 82}, 1406 (1999).
\bibitem{marte} A. Marte, {\cal et al.}, Phys.
Rev. Lett. {\bf 89}, 283202 (2002).
\bibitem{Loftus} T. Loftus, C. A. Regal, C. Ticknor, J. L. Bohn and D. S.
Jin, Phys. Rev. Lett. {\bf 88}, 173201 (2002).
\bibitem{Jochim}  S. Jochim, M. Bartenstein, G. Hendl, J. Denschlag, 
R. Grimm, A. Mosk and M. Weidem\"{u}ller
Phys. Rev. Lett. {\bf 89}, 273202 (2002).
\bibitem{Ohara} K. M. O'Hara, S. L. Hemmer, S. R. Granade, M. E. Gehm, 
J. E. Thomas, V. Venturi, E. Tiesinga, and C.J. Williams,
Phys. Rev. A. {\bf 66}, 
041401 (2002).
\bibitem{Dieckmann} K. Dieckmann, C. A. Stan, S. Gupta, Z. Hadzibabic, 
C.H. Schunck, and W. Ketterle, Phys. Rev. Lett. {\bf 89}, 203201 (2002).
\bibitem{Regal} C. A. Regal, C. Ticknor, J. L. Bohn, and D. S.
Jin, Phys. Rev. Lett. {\bf 90}, 053201 (2003).
\bibitem{Stoof} H.T.C. Stoof, M. Houbiers, C. A. Sackett, and R.G.
Hulet, Phys. Rev. Lett. {\bf 76}, 10 (1996).
\bibitem{Holland} M. Holland, S.J.J.M.F. Kokkelmans, M.L. Chiofalo,
and R. Walser, Phys. Rev. Lett. {\bf 87}, 120406 (2001).
\bibitem{Esry} H. Suno, and B. D. Esry, C. H. Greene,  Phys. Rev. Lett.
{\bf 90}, 53202 (2003).
\bibitem{Regal2} C. A. Regal and D. S. Jin, Phys. Rev. Lett. {\bf 90},
230404 (2003).
\bibitem{Bourdel} T. Bourdel, {\cal et al.}, Phys. Rev. Lett. 
{\bf 91}, 20402 (2003).
\bibitem{Gupta} S. Gupta, {\cal et al.}, Science {\bf300}: 1723-1726; 
\bibitem{Regal3} C. A. Regal, C. Ticknor, J. L. Bohn, and D. S.
Jin, Nature {\bf 424}, 47 (2003).
\bibitem{Strecker} K. E. Strecker, G. B. Partridge, R. G. Hulet. 
Phys. Rev. Lett. {\bf 91}, 80406 (2003);
\bibitem{Cub}J. Cubizolles {\cal et al.}, cond-mat 030801 (2003);
\bibitem{Jochim2} S. Jochim, {\cal et al.}, cond-mat 0308095 (2003);
\bibitem{Demarco} B. DeMarco, J. L. Bohn, J.P. Burke, Jr., M. Holland, and 
D. S. Jin, Phys. Rev. Lett. {\bf 82}, 4208 (1999).
\bibitem{polar} 
For a review see: 
M. Baranov, L. Dobrek, K. Goral, L. Santos, and M. Lewenstein, 
Physica Scripta, {\bf T 102}, 74 (2002).
\bibitem{Johnson} B.R. Johnson, J. Comp. Phys., {\bf 13}, 445 (1973).
\bibitem{Burke2} J. Burke, Ph.D. Thesis, University of Colorado (1999).
\bibitem{triplet}L. Li, A. M. Lyyra, W. T. Luh, and W. C. Stwalley, 
J. Chem. Phys. {\bf93}, 8452 (1990),         
W. T. Zemke,C. Tsai, and W. C. Stwalley, J. Chem. Phys. {\bf101}, 10382 (1994) 
\bibitem{singlet} C. Amiot J. Mol. Spectroscopy {\bf147}, 370 (1991).
\bibitem{Brink} D. M. Brink and G. R. Satchler, {\cal Angular Momentum}, 
Clardeon Press, Oxford \copyright 1993.
\bibitem{MM} N.F. Mott and H. S. W. Massey, {\cal The Theory of Atomic 
Collisions} 3$^{rd}$ Ed., Clardeon Press, Oxford \copyright 1965 
\bibitem{Roberts} J. L. Roberts, James P. Burke, Jr., N. R. Claussen, 
S. L. Cornish, E. A. Donley, and C. E. Wieman, Phys Rev. A 
{\bf 64}, 24702 (2001). 
\bibitem{Simoni} A. Simoni, F. Ferlaino, G. Roati, G. Modugno, and 
M. Inguscio, Phys. Rev. Lett. {\bf 90}, 163202 (2003).
\end{thebibliography}
\end{document}